\documentclass[twocolumn,amsmath,amssymb,prb,superscriptaddress]{revtex4-1}
\pdfoutput=1

\usepackage{bm}      
\usepackage{graphicx}

\begin{document}
\draft

\title{{\it Ab initio} calculation of $d$--$d$ excitations in quasi-one-dimensional Cu $d^9$ correlated materials}

\author{Hsiao-Yu Huang}
\affiliation{Institute for Theoretical Solid State Physics, IFW Dresden, Helmholtzstr.~20, 01069 Dresden, Germany}
\affiliation{College of Science, National Tsing Hua University, 30013 Hsinchu, Taiwan}
\affiliation{National Synchrotron Radiation Research Center, Hsin-Ann Rd.~101, 30076 Hsinchu, Taiwan}
\author{Nikolay A.~Bogdanov}
\affiliation{Institute for Theoretical Solid State Physics, IFW Dresden, Helmholtzstr.~20, 01069 Dresden, Germany}
\affiliation{National University of Science and Technology MISIS, Leninskiy~pr.~4, 119049 Moscow, Russia}
\author{Liudmila Siurakshina}
\affiliation{Institute for Theoretical Solid State Physics, IFW Dresden, Helmholtzstr.~20, 01069 Dresden, Germany}
\affiliation{Max-Planck-Institut f\"{u}r Physik komplexer Systeme, N\"{o}thnitzer Str.~38, 01187 Dresden, Germany}
\affiliation{Laboratory of Information Technologies, Joint Institute for Nuclear Research, 141980 Dubna, Russia}
\author{Peter Fulde}
\affiliation{Max-Planck-Institut f\"{u}r Physik komplexer Systeme, N\"{o}thnitzer Str.~38, 01187 Dresden, Germany}
\affiliation{POSTECH, San 31 Hyoja-dong, Namgu Pohang, Gyeongbuk 790-784, Korea}
\author{Jeroen van den Brink}
\author{Liviu Hozoi}
\affiliation{Institute for Theoretical Solid State Physics, IFW Dresden, Helmholtzstr.~20, 01069 Dresden, Germany}

\begin{abstract}
With wavefunction-based electronic-structure calculations we determine the Cu $d$--$d$ excitation
energies in quasi-one-dimensional spin-chain and ladder copper oxides.
A complete set of local excitations has been calculated for cuprates with corner-sharing
(Sr$_2$CuO$_3$ and SrCuO$_2$) and edge-sharing (LiVCuO$_4$, CuGeO$_3$, LiCu$_2$O$_2$ and Li$_2$CuO$_2$)
CuO$_4$ plaquettes, with corner-sharing CuF$_6$ octahedra (KCuF$_3$), for the ladder system
CaCu$_2$O$_3$, and for multiferroic cupric oxide CuO.
Our data compare well with available results of optical absorption measurements
on KCuF$_3$ and the excitation energies found by resonant inelastic x-ray scattering experiments
for CuO.
The {\it ab initio} results we report for the other materials should be helpful for the interpretation
of future resonant inelastic x-ray scattering experiments on those highly anisotropic compounds.
\end{abstract}

\date\today

%% \pacs{PACS numbers: 71.27.+a, 71.70.-d, 71.70.Gm, 74.70.-b}
\maketitle

\section{Introduction}

Many-body effects within the partially filled $3d$ shells of transition-metal oxides
are at the heart of various intriguing phenomena.
Much of the interesting physics comes from the interplay between electron delocalization
caused by intersite orbital overlap and strong Coulomb interactions within the $d$ shell.
The investigation of this interplay goes back as far as the 1930's with the work of Verwey,
Mott, and others.\cite{3d_oxides_1937}
Additional degrees of freedom arising from orbital degeneracy, electron-lattice couplings,
and/or ligand $p$ to metal $d$ charge transfer effects increase in many cases the complexity of
the problem.
Further, it was found that the multiorbital couplings within the $3d$ shell may be strong
even when the orbital degeneracy is lifted, e.g., by the electrostatic field set up by
neighboring ions.
For the copper oxide superconductors, for example, it is believed that the off-diagonal coupling
between states of $x^2\!-\!y^2$ and $z^2$ symmetry is one of the elements which determine the precise
shape of the Fermi surface, see, e.g.,
\onlinecite{CuO2_z2_eskes91,CuO2_z2Tc_ohta91,CuO2_z2_feiner96,CuO2_QPs_hozoi08,CuO2_z2Tc_sakaki10}
and references therein.
Accurate estimates for the size of the $d$-level splittings are therefore important in designing
realistic model Hamiltonians for these systems.

Traditionally the $d$--$d$ excitations in transition-metal oxides have been investigated by optical
spectroscopy.
First measurements on rocksalt compounds such as NiO and CoO were reported in the late 50's.
\cite{dd_NiO_CoO_59}
Although in centro-symmetric cases these transitions are not dipole allowed, they do aquire weak
intensity in the optical spectra due to the admixture of lattice vibrations which lift the
inversion symmetry at the transition-metal site.

It has been shown recently that highly accurate measurements for both magnetic 
and local charge excitations can be carried out by resonant inelastic x-ray scattering (RIXS).
\cite{RIXS_review_11}
As demonstrated by Moretti Sala {\it et al.} \cite{CuO2_rixsdd_11} for the case of $d$--$d$
transitions in two-dimensional (2D) Cu oxide systems, the high energy resolution and the analysis of
the polarization and scattering geometry dependence allow a quite reliable interpretation as
concerns the nature of the final excited states.

The calculation of local excitations within the transition-metal $3d$ shell is an interesting
problem.
It is desirable to use alternative approaches to standard density functional theory (DFT) calculations
because in first place electron correlations are strong and secondly, in its original formulation,
DFT is only a ground-state theory.
The {\it ab initio} wavefunction-based quantum chemical approaches therefore constitute the method 
of choice here.
Advanced quantum chemical calculations have been recently applied to the study of $d$--$d$
excitations in La$_2$CuO$_4$, Sr$_2$CuO$_2$Cl$_2$, and CaCuO$_2$.
The {\it ab initio} results\cite{dd_CuO2_hozoi_11} turned out to be in good agreement with the RIXS data reported by
Moretti Sala {\it et al.} \cite{CuO2_rixsdd_11}
On the other hand, recent DFT calculations on La$_2$CuO$_4$ predict $d_{x^2-y^2}$ to $d_{z^2}$
excitation energies that are 0.5 eV lower than in experiment.\cite{CuO2_z2Tc_sakaki10,CuO2_rixsdd_11}

Here, we extend the study of 2D cuprates as in Ref.~\onlinecite{dd_CuO2_hozoi_11} to the case of highly anisotropic
chain and ladder Cu $d^9$ compounds.
The motivation for the present investigation is twofold.
One one hand, we want to check the performance of our quantum chemical computational scheme\cite{dd_CuO2_hozoi_11}
in the case of one-dimensional (1D) and quasi-1D Cu $d^9$ systems.
Quantum chemical calculations on relatively small clusters have been earlier performed on both
2D and 1D cuprates, e.g., the 2D layered materials La$_2$CuO$_4$ and Sr$_2$CuO$_2$Cl$_2$ and the
chain-like system Sr$_2$CuO$_3$.\cite{CuO2_dd_coen00}
However, differences as large as 0.5 eV were found between the on-site $d$--$d$ excitation
energies reported for La$_2$CuO$_4$ and Sr$_2$CuO$_2$Cl$_2$ in Ref.~\onlinecite{CuO2_dd_coen00} and
our more recent results discussed in Ref.~\onlinecite{dd_CuO2_hozoi_11}.
Those differences seem to be related to the less precise description in Ref.~\onlinecite{CuO2_dd_coen00}
of the adjacent $3d$-metal and O ions, i.e., the nearest neighbor (NN) Cu ions and ligands
around the CuO$_4$ plaquette at which the $d$--$d$ excited states are explicitly computed.
Similar differences are here found for the $d$-level splittings of the chain system Sr$_2$CuO$_3$,
which shows that a careful analysis of this issue is indeed motivated.

Secondly, our {\it ab initio} data should be helpful for the correct interpretation of RIXS
and optical spectra in these compounds.
For highly anisotropic structures, the degeneracy of both the $t_{2g}$ and $e_g$ levels is lifted 
and the excitation spectra display a very rich structure.
Even in the 2D compounds, the sequence of the different excited states cannot always be predicted
beforehand.
For example, due to the different ratios between the in-plane and apical Cu--O bond lengths, the
$z^2$ hole state corresponds to the lowest crystal-field excited state in La$_2$CuO$_4$, with a relative
energy of 1.4 eV,\cite{dd_CuO2_hozoi_11,CuO2_rixsdd_11} and to the highest crystal-field excitation
in HgBa$_2$CuO$_4$, with a relative energy of 2.1 eV.\cite{dd_CuO2_hozoi_11}
The situation should be even more complex in 1D systems.
Reliable {\it ab initio} results are therefore desirable for the chain and ladder cuprates.

\section{Computational details}

Our computational approach is based on correlated {\it ab initio} methods traditionally
used in quantum chemical studies on molecular systems.
For each of the materials investigated here, the starting point is a restricted Hartree-Fock
(RHF) calculation with periodic boundary conditions.
All RHF calculations were performed with the {\sc crystal} program package.\cite{crystal}
We applied Gaussian-type atomic basis sets from the {\sc crystal} library, i.e.,
basis functions of triple-zeta quality\cite{QC_book_00} for Cu, O, and F and basis sets of either 
double-zeta or triple-zeta quality for metal ions next to the CuO$_2$ chains
(e.g., Li, Ca or V).
In all computations experimental lattice parameters were used, as reported in
Refs.~\onlinecite{Sr2CuO3, SrCuO2, CuO, CuGeO3, Li2CuO2, LiVCuO4, LiCu2O2, CaCu2O3, KCuF3}.

Post Hartree-Fock correlation calculations can be carried out on finite embedded clusters,
due to the local character of the correlation hole of a $d$ electron.
Yet, the orbitals used here are those of the infinite system.
Each embedded cluster $\mathcal{C}$ consists of two distinct regions:
an active region $\mathcal{C}_A$ where the actual correlation treatment is performed and
a buffer region $\mathcal{C}_B$ whose role is to provide support for the longer-range tails
of Wannier orbitals (WO's) centered at sites in $\mathcal{C}_A$.
The active region $\mathcal{C}_A$ includes one reference Cu site, the NN ligands, and
the NN Cu ions.
For the systems addressed in this study, a given Cu ion may have four, five or six NN
ligands.
Those NN ligands thus form either L$_4$ plaquettes, L$_5$ pyramids or distorted L$_6$
octahedra around a particular Cu site.
As concerns the $\mathcal{C}_B$ region, we include in there each ligand coordination cage
around the NN Cu sites and all NN closed-shell metal ions.
In CuGeO$_3$, for example, there are 8 Ge$^{4+}$ NN's.
In LiCu$_2$O$_2$, there are 5 Cu$^{1+}$ $3d^{10}$ and 8 Li$^{1+}$ NN's.
All post-RHF computations were performed with the {\sc molpro} quantum chemical software.
\cite{molpro_2006}

The orbital basis associated with a given cluster $\mathcal{C}$ is a set of projected
WO's: 
localized WO's associated with the RHF bands are first derived with the help of the
orbital localization module of the {\sc crystal} package and subsequently projected
onto the set of Gaussian basis functions associated with the atomic sites within the
cluster $\mathcal{C}$. 
Technicalities concerning this procedure are discussed in
Refs.~\onlinecite{dd_CuO2_hozoi_11,crystal_molpro_machinery_pks,QC_review_reimers11}.
For each particular cluster, the RHF data is additionally used to generate an
one-electron embedding potential that models the crystalline environment.
This effective potential is constructed with the {\sc matrop} module of the
{\sc molpro} program by using a real-space matrix representation of the 
self-consistent Fock operator from the periodic RHF calculation.
\cite{crystal_molpro_machinery_pks,QC_review_reimers11}
All necessary RHF data is converted into {\sc molpro} format with the help
of an interface program.\cite{crystal_molpro_int}
Although the WO's at the atomic sites of $\mathcal{C}$ are derived for each of the
compounds discussed here by periodic RHF calculations for the Cu $3d^9$ electron
configuration, the embedding potentials are obtained by replacing the Cu$^{2+}$ $3d^9$
ions by closed-shell Zn$^{2+}$ $3d^{10}$ species.
This is a good approximation for the farther $3d$-metal sites, as the comparison between
our results and RIXS data shows.\cite{CuO2_rixsdd_11,dd_CuO2_hozoi_11}
The extension of this embedding scheme toward the construction of open-shell embeddings
is an ongoing project in our group.

While the occupied WO's in the buffer region $\mathcal{C}_B$ are kept frozen, all
valence orbitals centered at ligand and Cu sites in $\mathcal{C}_A$ are further
reoptimized\cite{dd_CuO2_hozoi_11} in multiconfiguration complete-active-space self-consistent-field 
(CASSCF) calculations.\cite{QC_book_00}
In the latter, the ground-state wavefunction and the crystal-field excited
states at the central Cu site are computed by state-averaged multiroot optimizations.
\cite{QC_book_00}
The Cu $d$-level splittings are finally obtained from additional multireference single 
and double configuration-interaction\cite{QC_book_00} (MRCI) calculations as the relative energies of
the crystal-field excited states.
The central Cu $3s$, $3p$, $3d$, NN ligand $2p$, and NN half-filled Cu $3d$ orbitals
are correlated in MRCI. 

The virtual orbital space in the MRCI calculation cannot be presently restricted
just to the $\mathcal{C}_A$ region.
It thus includes virtual orbitals in both $\mathcal{C}_A$ and $\mathcal{C}_B$, which
leads to very large MRCI expansions.
To make the computations feasible, we restrict our study to ferromagnetic (FM) allignment
of the Cu $d$ spins. 
Even for FM clusters, the MRCI expansion may include in some cases up to $\sim\!10^{11}$
Slater determinants.

\section{Results and discussions}

\subsection{Chains of corner-sharing CuO$_4$ plaquettes}

The 1D compounds Sr$_{2}$CuO$_{3}$ and SrCuO$_{2}$, built of chains of
corner-sharing CuO$_4$ plaquettes, display a number of quite unusual properties.
On one hand, these materials have the largest NN antiferromagnetic (AF) exchange integrals in
the family of Cu $d^9$ oxides.
Yet, the interchain couplings are very weak.
\cite{s2co3_J_suzuura96,s2co3_J_eder97,s2co3_J_coen00,s2co3_sco2_J_motoy96,sco2_J_zaliz04, s2co3_J_Rosner97}
They thus constitute ideal systems for studying the magnetic response of 1D spin-1/2
antiferromagnets.
Secondly, high-resolution angle-resolved photoemission experiments on these compounds have
for the first time revealed the realization of spin-charge separation in 1D electron systems.
\cite{spin_charge_97_99,spin_charge_Kim06}

%% FIGURE
\begin{figure}[!b]
\includegraphics[width=0.99\columnwidth]{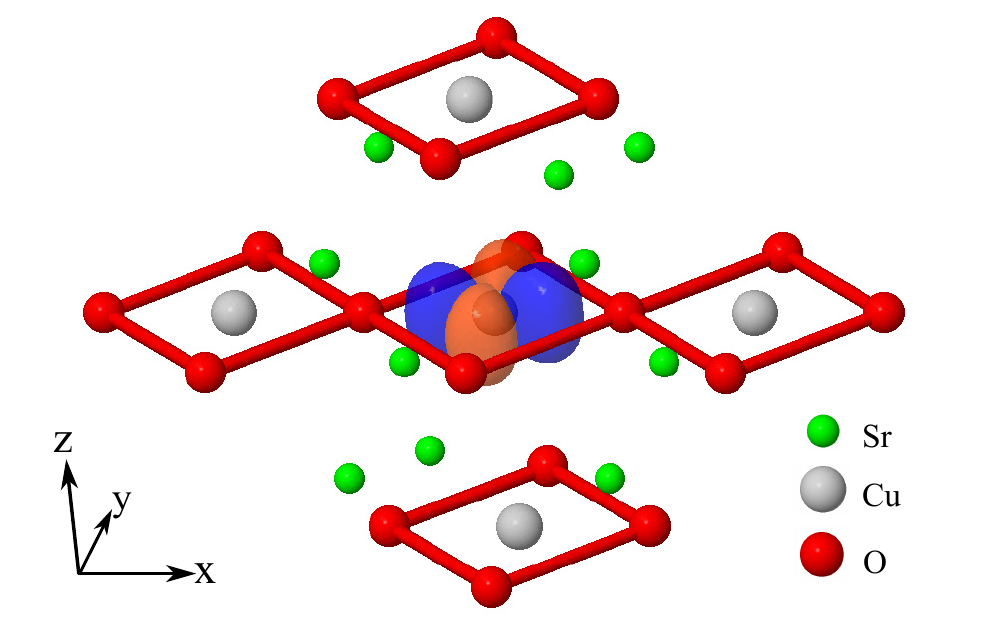}
\caption{
Chain of corner-sharing CuO$_4$ plaquettes in Sr$_2$CuO$_3$.
The ground-state hole orbital is $d_{x^2-y^2}$.
}
\label{sr2cuo3}
\end{figure}

We list in Table \ref{dd_corner} results for the $d$-level splittings of Sr$_{2}$CuO$_{3}$
and SrCuO$_{2}$.
As concerns the crystallographic details, a major difference between these two compounds
is the presence of a network of single CuO$_2$ chains in Sr$_{2}$CuO$_{3}$,\cite{Sr2CuO3}
see Fig.~1, whereas SrCuO$_{2}$ displays a double-chain structure, with CuO$_4$ plaquettes on adjacent
chains sharing common edges. \cite{SrCuO2}
The clusters we use in our CASSCF and MRCI calculations include therefore two NN plaquettes
in the case of Sr$_{2}$CuO$_{3}$ and four adjacent plaquettes for SrCuO$_{2}$.
The reference system is chosen such that the $x$-axis is along the CuO$_2$ chains
and $z$ is perpendicular onto the CuO$_4$ plaquettes.
In this framework, the ground-state hole orbital has $x^{2}\!-\!y^{2}$ symmetry.
The data in Table \ref{dd_corner} show that the $d$-level splittings in Sr$_{2}$CuO$_{3}$
and SrCuO$_{2}$ are very similar.
In both cases, the relative energy of each particular excited state is not very different
from the corresponding value in the Li$_{2}$CuO$_{2}$ compound, see Sec.~III.C.
As in Sr$_{2}$CuO$_{3}$ and SrCuO$_{2}$, there are no apical ligands in Li$_{2}$CuO$_{2}$
either.
Results for the $d$-level electronic structure of the latter compound are listed in
Table \ref{dd_edge}.

Inclusion of single and double excitations on top of the CASSCF wavefunctions brings
a nearly uniform upward shift of 0.1 to 0.2 eV for the $d$--$d$ transitions.
By adding Davidson corrections,\cite{footnote_Davidson} the relative energies of the excited states
further increase by 0.05 to 0.1 eV.
The MRCI values listed in Table \ref{dd_edge} include such Davidson correction
terms.
This relative stabilization of the electronic ground state with respect to the higher
lying states in MRCI is mainly related to ligand $2p$ to metal $3d$ charge-transfer correlation
effects which are more important for the in-plane $d_{x^2-y^2}$ hole orbital having $\sigma$-type
overlap with the NN O $2p$ functions.

%% TABLE 1
\begin{table}[!t]
\caption{
CASSCF/MRCI $d$--$d$ excitation energies for corner-sharing chains of CuO$_4$ plaquettes
in Sr$_{2}$CuO$_{3}$ and SrCuO$_{2}$ (eV).
The MRCI values include Davidson corrections.\cite{footnote_Davidson}
}
\begin{ruledtabular}
\begin{tabular}{llll}
Hole orbital &Sr$_{2}$CuO$_{3}$ &SrCuO$_{2}$ \\
 &CASSCF/MRCI &CASSCF/MRCI\\
\hline
$x^{2}-y^{2}$ &$0$ &$0$ \\
$xy$ &$1.20$/$1.50$ &$1.26$/$1.55$ \\
$yz$ &$1.83$/$2.14$ &$1.77$/$2.03$ \\
$xz$ &$1.90$/$2.24$ &$1.88$/$2.19$ \\
$z^{2}$ &$2.16$/$2.55$ &$2.08$/$2.44$ \\
\end{tabular}
\end{ruledtabular}
\label{dd_corner}
\end{table}

As concerns the comparison with other theoretical investigations,
there are differences of 0.3 eV or more between the $d$--$d$ excitation
energies listed for Sr$_{2}$CuO$_{3}$ in Table \ref{dd_corner} and the values computed
earlier in Ref.~\onlinecite{CuO2_dd_coen00}.
We presume that these large differences, e.g., about 0.5 eV for the $yz$ hole states, are mainly related
to the approximations used in Ref.~\onlinecite{CuO2_dd_coen00} for representing the Cu
and O ions on NN plaquettes.
While here we represent those species at the all-electron level, in the earlier study
\cite{CuO2_dd_coen00} the NN Cu and O ions were modeled by Mg$^{2+}$ ions and formal
$2-$ point charges, respectively.
Point charges were also used for the embedding in Ref.~\onlinecite{CuO2_dd_coen00}.

We note at this point that in an AF lattice the total energy of a given
state within the $d^n$ manifold is a sum of a crystal-field contribution, i.e., an
on-site crystal-field splitting, and a magnetic term (see also the discussion in
Refs.~\onlinecite{RIXS_review_11,dd_CuO2_hozoi_11}).
As mentioned above, the AF NN spin coupling constant $J$ is remarkably large in
Sr$_{2}$CuO$_{3}$ and SrCuO$_{2}$, 0.20 to 0.25 eV
\cite{s2co3_J_suzuura96,s2co3_J_eder97,s2co3_J_coen00,s2co3_sco2_J_motoy96,sco2_J_zaliz04}.
From the exact Bethe-ansatz solution for the 1D Heisenberg Hamiltonian,
\cite{1D_AF_bethe,1D_AF_cloizeaux,1D_AF_fadeev81} the AF ground-state stabilization
energy is $J\ln2$.
On the other hand, from overlap considerations, we conclude that for the crystal-field
excited states the (super)exchange with the NN Cu $d_{x^2-y^2}$ spins is either zero or
much weaker.
For technical reasons, see the discussion in Sec.~II, the quantum chemical calculations
were here performed for a FM cluster.
For a meaningful comparison between the MRCI results and experimental RIXS data, one
should therefore subtract a term $J\ln2$ from the relative RIXS energies, representing
the energy stabilization of the AF ground-state with respect to the crystal-field
excited states.
These considerations are relevant in light of future RIXS measurements on AF 1D
cuprates.

%% TABLE 2
\begin{table}[!t]
\caption{
Cu $d$--$d$ excitation energies in KCuF$_3$ (eV).
The ground-state Cu $t_{2g}^6d_{y^2}^2d_{x^2-z^2}^{1}$ configuration is taken as reference.
The Jahn-Teller distortions occur within the $xy$ plane.
The MRCI values include Davidson corrections.\cite{footnote_Davidson}
}
\begin{ruledtabular}
\begin{tabular}{llll}
Hole orbital &CASSCF &MRCI &Experiment$^a$\\
\hline
$x^2-z^2$ &$0$ &$0$ &$0$ \\
$y^2$ &$0.76$ &$0.85$ &$0.71-1.02$ \\
$xz$ &$0.89$ &$1.01$ &$1.05-1.15$ \\
$xy$ &$1.04$ &$1.17$ &$1.21-1.37$ \\
$yz$ &$1.11$ &$1.25$ &$1.34-1.46$ \\
\end{tabular}
\end{ruledtabular}
$^a$: Optical absorption, Ref.~\onlinecite{KCuF3_dd_optics08}.
The numbers correspond to the onsets and the maxima of the absorption bands.
\end{table}

\subsection{Corner-sharing CuF$_6$ octahedra in KCuF$_{3}$}

KCuF$_{3}$ is a prototype material for systems with strong coupling among the charge,
orbital, and spin degrees of freedom.\cite{KCuF3_kadota_67,OO_kugel_khomskii_82}
The crystalline structure of this compound is perovskite-like,\cite{KCuF3} i.e.,
three-dimensional (3D).
The degeneracy of the Cu $e_g$ levels in an ideal perovskite lattice is lifted however
through cooperative Jahn-Teller distortions.
The latter imply a configuration with alternating, longer and shorter, Cu--F bonds for
Cu ions along the $x$ and $y$ axes and orbital ordering in the $xy$ plane.
The hole in the Cu $3d$ shell thus alternately occupies $3d_{x^2-z^2}$ and $3d_{y^2-z^2}$
orbitals.

In the $xy$ plane, the magnetic couplings are weak and FM.
On the other hand, along the $z$-axis the NN $J$ is large and AF.
\cite{KCuF3_j_tennant95}
Actually this makes KCuF$_{3}$ a close to ideal 1D Heisenberg antiferromagnet.
The predictions for the low-lying spin excitations of the 1D AF Heisenberg chain
\cite{1D_AF_cloizeaux,1D_AF_fadeev81} have been confirmed by inelastic neutron
scattering experiments.\cite{KCuF3_AF1D_tennant93}

The magnetic behavior changes from 1D to 3D at the N\'{e}el temperature $T_{\rm N}\!=\!39$
K.
The emergence of sharp crystal-field absorption peaks in the optical spectra approximately
10 K above $T_{\rm N}$ has been interpreted as evidence for a symmetry change related 
with a crossover from dynamic to static displacements of the F ions.
\cite{KCuF3_dd_optics08}
In this picture, the orbital and 3D AF ordering are intimately related, in the sense
that the former paves the road for the latter.

The $d$--$d$ excitation energies seen in the optical spectra were compared in
Ref.~\onlinecite{KCuF3_dd_optics08} with the outcome of DFT band-structure
calculations.
However, strictly speaking, DFT is a ground-state theory.
Here, we provide results of excited-state CASSCF and MRCI calculations for the Cu
$d$-level splittings in KCuF$_{3}$, see Table V.
The MRCI treatment brings corrections of 0.1--0.15 eV to the CASSCF splittings, somewhat
smaller than for the oxides discussed above.
This is related to the more ionic character of the Cu--F bond in KCuF$_{3}$ and
smaller degree of Cu $3d$ and F $2p$ orbital mixing. 
As a general trend, the MRCI results tend to slightly underestimate the values
corresponding to the maxima of the experimental absorption peaks, by 0.1--0.2 eV.
%% Even better agreement with the experiment would be achieved by using larger Cu and F basis
%% sets in the quantum chemical calculations.
One obvious effect of the less anisotropic environment is a much smaller splitting
between the two $e_g$ components.
This particular electronic-structure parameter is in fact the smallest among the Cu
$d^9$ compounds investigated here.

\subsection{Chains of edge-sharing CuO$_4$ plaquettes}

%% FIGURE
\begin{figure}[!b]
\includegraphics[width=0.99\columnwidth]{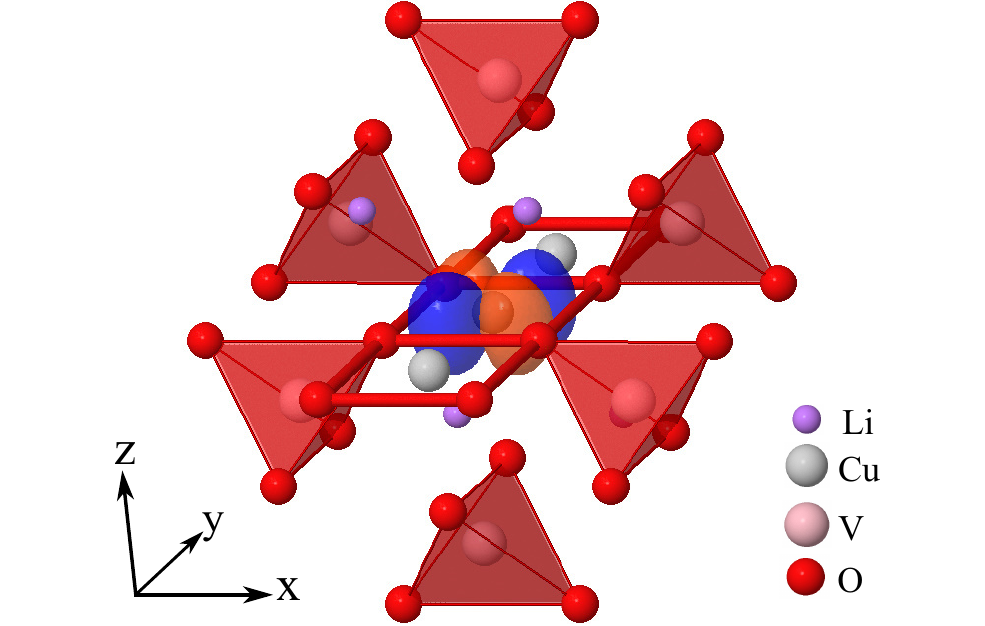}
\caption{
Chain of edge-sharing CuO$_{4}$ plaquettes in LiVCuO$_{4}$.
The ground-state hole orbital is $d_{xy}$ in the chosen reference system.
The six NN VO$_4$ octahedra around the reference Cu site are highlighted in the figure.
}
\label{livcuo4}
\end{figure}

Cuprates in which the CuO$_4$ plaquettes are edge-sharing, are characterized by weak
superexchange interactions between NN Cu spins, which is due to Cu--O--Cu bond angles that
are close to perpendicular.
Depending on the detailed crystal structure, the NN interaction is usually FM
(LiVCuO$_{4}$, Li$_2$CuO$_{2}$)\cite{Enderle10,Drechsler11,Nishimoto11} or even AF (CuGeO$_3$).
\cite{Hase93,Hirata94}
The in-chain next-NN exchange is always AF and causes frustration independently of the
sign and size of the NN coupling.
Further, the longer-range intrachain and interchain couplings are sizeable as well in
some of these compounds and lead to an intricate competition of magnetic interactions.
This has resulted in an appreciable scientific interest in the edge-sharing chain systems.

In LiVCuO$_{4}$ the ratio of the FM NN coupling and the AF next-NN exchange --
the frustration parameter -- is controversial.\cite{Enderle10,Drechsler11}
For LiCu$_2$O$_2$ even the sign of the NN coupling constant gave rise to debate.
\cite{Matsuda05,Gippius04,Drechsler05}
It has been also argued that a large fourth-nearest-neighbor AF coupling constant is responsible
for the  incommensurate helimagnetic ground state of LiCu$_2$O$_2$.
\cite{Matsuda05}
In CuGeO$_3$ the NN exchange is AF and this system is famous to undergo a so-called
spin-Peierls transition\cite{Hase93,Hirata94} to an AF state with a gapped excitation spectrum.
When deriving the values of the shorter- and longer-range exchange interactions in terms of
downfolding techniques, also the $3d$-level excitation energies come into play.

%% TABLE x
\begin{table}[!t]
\caption{
CASSCF/MRCI $d$--$d$ excitation energies for edge-sharing chains of CuO$_4$ plaquettes
(eV).
The MRCI values include Davidson corrections.\cite{footnote_Davidson}
The $y$-axis is parallel to the CuO$_2$ chains, $z$ is perpendicular onto the
CuO$_4$ plaquettes, see text.
}
\begin{ruledtabular}
\begin{tabular}{lllll}
Hole orbital &LiVCuO$_{4}$ &CuGeO$_{3}$ &LiCu$_{2}$O$_{2}$ &Li$_{2}$CuO$_{2}$ \\
\hline
$xy$ &$0$ &$0$ &$0$ &$0$ \\
$x^{2}-y^{2}$ &$1.08$/$1.28$ &$1.19$/$1.41$ &$1.13$/$1.43$ &$1.20$/$1.52$ \\
$xz$ &$1.25$/$1.47$ &$1.42$/$1.61$ &$1.58$/$1.88$ &$1.72$/$2.04$ \\
$yz$ &$1.29$/$1.52$ &$1.45$/$1.64$ &$1.64$/$1.94$ &$1.72$/$2.04$ \\
$z^{2}$ &$0.98$/$1.18$ &$1.50$/$1.71$ &$1.67$/$1.98$ &$1.93$/$2.30$ \\
\end{tabular}
\end{ruledtabular}
\label{dd_edge}
\end{table}

The calculated CASSCF and MRCI results for the Cu $d$-level splittings in LiVCuO$_{4}$,
CuGeO$_{3}$, LiCu$_{2}$O$_{2}$, and Li$_{2}$CuO$_{2}$ are listed in Table \ref{dd_edge}.
The finite embedded clusters on which the correlation treatment is carried out
include one central and two NN CuO$_4$ plaquettes for the single chain compounds
LiVCuO$_{4}$, CuGeO$_{3}$, and Li$_{2}$CuO$_{2}$.
For the zigzag double-chain system LiCu$_{2}$O$_{2}$,\cite{LiCu2O2} the cluster $\mathcal{C}$ includes four
adjacent plaquettes.
The coordinate framework is chosen such that the $y$-axis is parallel to the CuO$_2$ chains
and $z$ is perpendicular onto a given CuO$_4$ plaquette.
In other words, there is a rotation of $45^{\circ}$ around $z$ as compared to the standard
coordinate framework for planar or octahedral coordination and the $x$ and $y$ axes 
are not along the in-plane Cu--O bonds.
The in-plane $\sigma$-type Cu $3d$ orbital that is singly occupied in the ground-state
configuration is thus denoted $d_{xy}$.

As for the corner-sharing plaquettes in Sr$_{2}$CuO$_{3}$ and SrCuO$_{2}$ (see Sec.~III.A),
inclusion of single and double excitations on top of the CASSCF wavefunctions brings 
a nearly uniform upward shift of 0.1 to 0.2 eV for the $d$--$d$ transitions.
By adding Davidson corrections, the relative energies of the excited states
further increase by 0.05 to 0.1 eV.
The MRCI values listed in Table \ref{dd_edge} include such Davidson correction
terms.
A second effect which deserves attention is the large variations found for the 
relative energy of the $d_{z^2}$ excited hole state.
It is known that the length of the apical Cu--O bond varies widely in Cu $d^9$
oxides.
The strong influence of the apical bond length on the $d_{z^2}$ hole excitation
was previously stressed for 2D cuprates in, e.g.,
Refs.~\onlinecite{CuO2_rixsdd_11,dd_CuO2_hozoi_11}.
Using simple electrostatic arguments, when the negative apical ions are closer
to the Cu site, less energy is required for exciting the in-plane Cu $3d$ hole
to the $d_{z^2}$ orbital pointing toward those apical ligands.
As concerns the chain-like compounds from Table \ref{dd_edge}, there are two 
apical O ions in CuGeO$_{3}$, one apical in LiCu$_{2}$O$_{2}$, and no apical 
ligand in Li$_{2}$CuO$_{2}$.
The relative energy of the $d_{z^2}$ hole state consequently increases from 1.71 eV
in CuGeO$_{3}$ to 1.98 in LiCu$_{2}$O$_{2}$ and to 2.30 eV in Li$_{2}$CuO$_{2}$.
In LiVCuO$_{4}$ there are two apical O ions as in CuGeO$_{3}$ but the
apical bond lengths are much shorter as compared to CuGeO$_{3}$, 2.49 versus 2.76
\AA \ (see Refs.~\onlinecite{CuGeO3,LiVCuO4}).
The $d_{xy}$ to $d_{z^2}$ excitation energy in LiVCuO$_{4}$ is thus substantially lower,
1.18 eV, even lower than in La$_2$CuO$_4$, where the apical bond length is
2.40 \AA \ and the transition to the $d_{z^2}$ orbital occurs at about 1.4 eV.
\cite{CuO2_rixsdd_11,dd_CuO2_hozoi_11}
However, one additional effect which comes into play in LiVCuO$_{4}$ is the
strong covalency between the V ions and both in-plane and apical O neighbors
of the Cu sites.
Such covalency effects on the VO$_4$ tetrahedra, see Fig.~2, give rise to large deviations
from the formal V$^{5+}$ and O$^{2-}$ valence states employed in a fully
ionic picture.
When large deviations from the formal value of $2-$ occur for the effective 
charges of {\it all} adjacent O ions, the strongest affected is actually the
ground-state energy because there are four O neighbors to which the lobes of
the Cu $d_{xy}$ orbital are directed.
%% and the in-plane 
%% bond lengths are much shorter than the apical bond lengths.
This ``destabilization'' of the ground-state energy of LiVCuO$_{4}$ explains the
nearly uniform downward shift of the $d_{x^2-y^2}$, $d_{yz}$, and $d_{xz}$
excited hole states, by about 0.15 eV as compared to CuGeO$_{3}$ (see Table
\ref{dd_edge}).

\subsection{CuO}

%% FIGURE
\begin{figure}[!b]
\includegraphics[width=0.99\columnwidth]{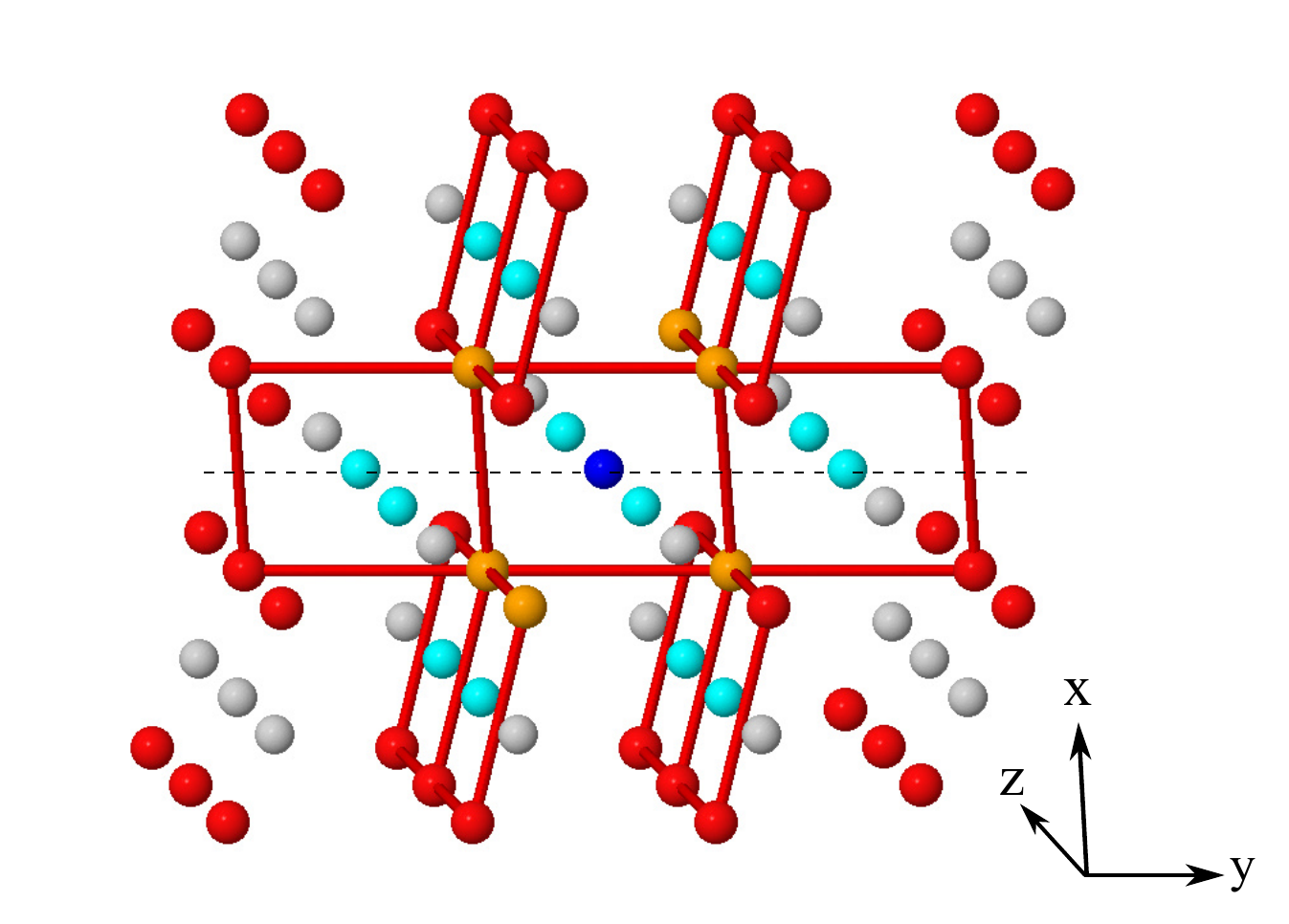}
\caption{
Crystal structure of CuO and sketch of the cluster used for the calculations.
The reference Cu and the fourteen adjacent Cu sites are shown in dark and
light blue, respectively.
O ions of the central octahedron are depicted in orange.
The thin dashed line parallel to the $y$-axis connects the central Cu site and
two Cu NN's along the same chain of edge-sharing CuO$_4$ plaquettes.
For each apical O, there are two NN Cu ions on adjacent chains parallel to the $y$-axis.
}
\label{cuo}
\end{figure}

In spite of its apparent simple chemical formula cupric oxide, CuO, is a complex material.
It was for instance only very recently discovered that it develops at the AF transition
temperature $T_{\rm N}$ of 230 K also a net ferroelectric polarization, which is of
substantial experimental\cite{Kimura08} and theoretical\cite{Mostovoy08,Giovannetti11}
interest.
As the ferroelectricity is completely due to magnetism, CuO is classified as a type-II
multiferroic.\cite{Brink08}
The complexity of this material system is related to the fact that each of the distorted
CuO$_6$ octahedra shares corners and edges with fourteen other, adjacent octahedra.
As for most of the Cu oxides, on a given octahedron the two apical-like Cu--O bonds
are much longer than the other four Cu--O bonds, 2.78 versus 1.95 or 1.96 \AA .
\cite{CuO}
The quantum chemical calculations were therefore carried out on a cluster whose 
active region $\mathcal{C}_A$ is defined by one reference CuO$_4$ plaquette and 
the nearest fourteen Cu neighbors, see Fig.~3.
The buffer region $\mathcal{C}_B$ includes the apical O ions of the reference Cu
site and for each adjacent Cu ion the nearest four ligands.

CASSCF and MRCI results for the Cu $d$-level splittings in CuO are given in Table
\ref{dd_CuO}.
CASSCF calculations were first performed for a FM configuration of the fifteen
Cu $d^9$ sites.
A non-standard local reference system was chosen for the central Cu site, with a
rotation of $45^{\circ}$ around $z$ as for the cuprates discussed in Sec.~III.C. 
The ground-state hole orbital is $d_{xy}$ in this framework.
CASSCF relative energies for the crystal-field excited states are listed in the
second column of Table \ref{dd_CuO}.

In a next step, MRCI calculations were carried out.
Due to the large number of open-shell NN Cu sites, for CuO we were forced to
restrict the CAS orbital reference space to the set of $3d$ orbitals of the central Cu ion.
Two different types of approximations were employed.
In a first set of restricted multiconfiguration calculations, out of the fourteen linear 
combinations of NN Cu $d_{xy}$ functions, we froze the occupation of the seven
lowest-energy (bonding-like) orbitals to 2 while the occupation of the higher-lying
(antibonding-like) orbitals was restricted to 0.
Only the Cu $3s$, $3p$, $3d$ and O $2p$ electrons on the central CuO$_4$ plaquette
were correlated in the subsequent MRCI treatment.
The difference between the CASSCF and MRCI values for each particular state in this
set of CASSCF and MRCI calculations was added afterwards to the initial CASSCF $d$--$d$
splitting.
The resulting numbers are listed in the third column of Table \ref{dd_CuO}.
In the second set of MRCI calculations, we replaced the Cu$^{2+}$ $d^9$ neighbors by 
closed-shell Zn$^{2+}$ $d^{10}$ ions.
Those MRCI correlation induced corrections were again added to the initial CASSCF
$d$--$d$ excitation energies and the corresponding results are given in the fourth
column of Table \ref{dd_CuO}.
As shown in Table \ref{dd_CuO}, the MRCI treatment brings shifts of 0.2 to 0.25 eV 
to the CASSCF splittings. 
In the two approximation schemes, the MRCI corrections for each particular state are
nearly the same, i.e., the MRCI correlation energies do not depend on details of
the NN $3d$ electron configuration.

%% TABLE x
\begin{table}[!t]
\caption{
CASSCF and MRCI $d$--$d$ excitation energies for CuO, see text.
The MRCI values include Davidson corrections.\cite{footnote_Davidson}
All numbers are in eV.
}
\begin{ruledtabular}
\begin{tabular}{lllll}
Hole orbital &CASSCF &MRCI-1 &MRCI-2 \\
\hline
$xy$ &$0$ &$0$ &$0$\\
$x^{2}-y^{2}$ &$1.13$ &$1.37$ &$1.38$\\
$yz$ &$1.54$ &$1.75$ &$1.79$\\
$xz$ &$1.60$ &$1.81$ &$1.85$\\
$z^{2}$ &$1.71$ &$1.93$ &$1.96$\\
\end{tabular}
\end{ruledtabular}
\label{dd_CuO}
\end{table}

The lowest crystal-field excitation is from $d_{xy}$ to $d_{x^2-y^2}$ and estimated
in Table \ref{dd_CuO} at about 1.4 eV.
This value compares well with the measured optical gap, 1.3--1.6 eV.
\cite{Kottyberg_1982,Ghijsen_1988,Marabelli_1995,Ito_1998}
RIXS measurements on CuO by Ghiringhelli \textit{et al.} \cite{Ghiringhelli_2009} show
that most of the weight of the $d$--$d$ excitation spectrum is between 1.7 and 2.3 eV,
with clear peaks at 1.85 and 2.15 eV.
Our MRCI data seem to somewhat underestimate those RIXS results, even when a $J\ln2$
correction is considered to account for the superexchange interactions along the
AF chains in CuO.\cite{footnote_J_CuO}
More flexible Cu and O basis sets are expected to further reduce these small differences
between theory and experiment.

%% FIGURE
\begin{figure*}[!t] 
\includegraphics[width=0.8 \textwidth]{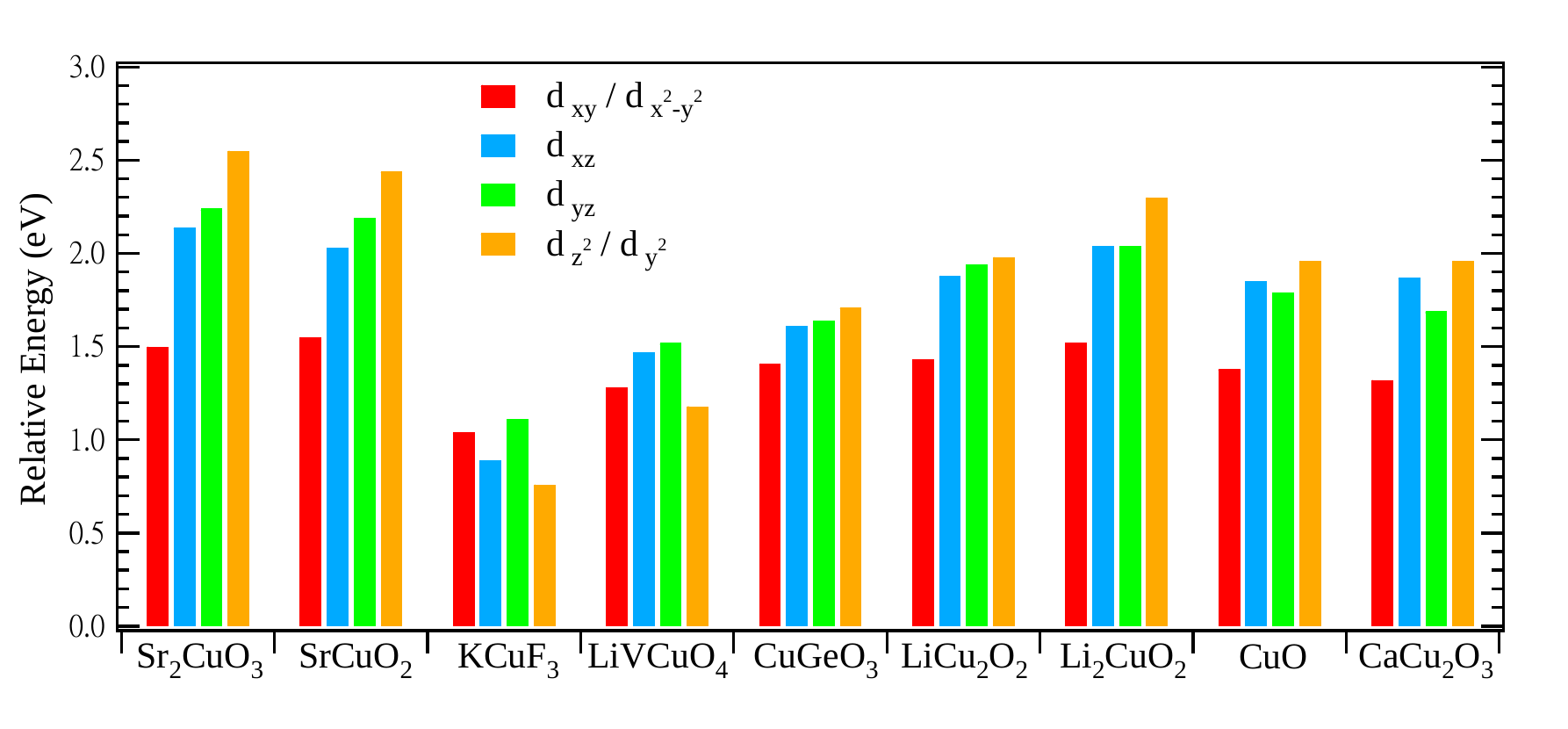}
\caption{
MRCI Cu $d$--$d$ excitation energies in 1D Cu $d^9$ compounds.
Depending on the reference coordinate system, see text, the symmetry of the lowest
$t_{2g}$ hole state (in red) is denoted either as $d_{xy}$ (corner-sharing chains
of plaquettes or octahedra) or $d_{x^{2}-y^{2}}$ (edge-sharing chains).
For KCuF$_3$, the $e_g$ excited state (in orange) has $y^2$ symmetry.
For all other compounds, the $e_g$ excited state has $z^2$ symmetry. 
} 
\label{bar} 
\end{figure*}

\subsection{A ladder cuprate: CaCu$_{2}$O$_{3}$}

In the so-called ladder compounds, the transition-metal ions are arranged
in a planar network of ladders which can display two or more legs.
These systems became subject of intense study when Dagotto {\it et al.}
\cite{Sladders_dagotto_9293} found theoretical evidence that the isolated 
$S\!=\!1/2$ AF two-leg ladder has a finite spin gap, i.e., a finite energy 
is needed to create a spin excitation.

Physical realizations are found in the vanadium oxide compound CaV$_2$O$_5$
\cite{CaV2O5_onoda96_iwase96} and in cuprates such as CaCu$_2$O$_3$, SrCu$_2$O$_3$,
\cite{Sladders_azuma94} and (Sr,Ca)$_{14}$Cu$_{24}$O$_{41}$.\cite{Sladders_srca14cu24o41}
The latter compound also displays superconductivity.\cite{Sladders_uehara96}

We analyze in this section the $d$-level electronic structure of CaCu$_2$O$_3$.
For most of the two-leg ladder cuprates, including CaCu$_2$O$_3$, octahedra
on the same ladder share common corners while adjacent octahedra on different
ladders share common edges.
One peculiar feature of CaCu$_2$O$_3$ is that on each rung of a ladder the
Cu--O--Cu angle is 123$^\circ$, much smaller than in other ladder compounds.\cite{CaCu2O3}
This is the reason the intersite magnetic couplings across the rungs are AF and small,
about 10 meV, while the superexchange along the leg of the ladder is AF and one 
order of magnitude larger, about 135 meV.\cite{Bordas_2005}
In other words, from the magnetic point of view the system is rather 1D, with
weakly interacting AF CuO$_2$ chains.
Choosing the $a$-axis parallel to the rung and $b$ along the leg of the ladder,
the ground-state hole orbital is $d_{x^2-y^2}$.

%% TABLE x
\begin{table}[!b]
\caption{
CASSCF and MRCI $d$--$d$ excitation energies for the ladder system
CaCu$_2$O$_3$ (eV).
The MRCI values include Davidson corrections.\cite{footnote_Davidson}
}
\begin{ruledtabular}
\begin{tabular}{lll}
Hole orbital &CASSCF &MRCI \\
\hline
$x^{2}-y^{2}$ &$0$ &$0$\\
$xy$ &$1.09$ &$1.32$\\
$yz$ &$1.46$ &$1.69$\\
$xz$ &$1.61$ &$1.87$\\
$z^{2}$ &$1.67$ &$1.96$\\
\end{tabular}
\end{ruledtabular}
\label{dd_ladder}
\end{table}

The quantum chemical calculations were carried out on a cluster whose active
region $\mathcal{C}_A$ is defined by one reference CuO$_4$ plaquette and the five
adjacent Cu ions (one Cu NN on the same rung, two leg NN's, and two Cu NN's 
on an adjacent ladder).
The buffer region $\mathcal{C}_B$ includes the two apical O ions of the reference
Cu site, the nearest four ligands for each adjacent Cu ion, and six Ca neighbors.
The apical O ligands are at rather large distance from the central Cu site, 
3.0~\AA .\cite{CaCu2O3}

CASSCF and MRCI results for the Cu $d$-level splittings in CaCu$_2$O$_3$ are listed 
in Table~\ref{dd_ladder}.
The MRCI treatment and the Davidson corrections bring shifts of 0.2 to 0.3 eV to the CASSCF
$d$--$d$ excitation energies.
The excitation energies to the $xz$, $yz$, and $z^2$ orbitals are somewhat smaller than
in Sr$_{2}$CuO$_{3}$, SrCuO$_{2}$, and Li$_2$CuO$_{2}$ because in the latter compounds there
are no apical ligands.
At the same time, the splitting between $d_{x^2-y^2}$ and $d_{z^2}$ is larger than in 
the other cuprates investigated here, see Fig.~4, because the apical Cu--O bond length is 
the largest in CaCu$_2$O$_3$.

\section{Conclusions}

It has been shown here that wavefunction-based electronic-structure calculations are
well-suited for computing local charge excitations as observed in transition-metal oxides 
by optical and RIXS measurements.
While this was demonstrated previously for a number of layered cuprates,\cite{dd_CuO2_hozoi_11} we
have extended here those applications to 1D and ladder compounds. %%with similar success.
Results and trends for the $d$--$d$ excitations in low-dimensional cuprates are summarized
in Fig.~4.
The {\it ab initio} data compare well with the results of optical absorption measurements
on KCuF$_3$\cite{KCuF3_dd_optics08} and somewhat underestimate the excitation energies
found by RIXS experiments for CuO.\cite{Ghiringhelli_2009}
Future RIXS experiments on other 1D cuprates included in Fig.~4 will constitute a direct
test for the symmetry and energy of the different $d$--$d$ excitations that we predict
here.

From a more general perspective, the results reported here clearly indicate the significant potential
of the {\it ab initio} wavefunction-based methods in the context of correlated electronic materials and the need to
further develop such techniques in parallel with approaches based on density functional theory.
The latter has revolutionized the field of electronic-structure calculations but its limits
are also clear.
Computations of excited states of strongly correlated electrons are in a way the worst case for
DFT.
One is then forced to avail oneself to alternatives, i.e., wavefunction-based methods.
An attractive feature is that strong correlations can be handled by CASSCF calculations even
when we deal with infinite periodic systems.
In all approximations which are being made, the accuracy of the wavefunction-based methods can be
progressively increased.
Here we supplemented the CASSCF calculations by MRCI, yet, alternative supplementary methods
are possible.
Further exploration of such {\it ab initio} techniques is planned for the future.

\section*{Acknowledgments}

We thank V.~Bisogni, K.~Wohlfeld, S.-L.~Drechsler, and D.~J.~Huang for fruitful discussions.
H.-Y.~H., N.~A.~B., and L.~H. acknowledge financial support from 
the National Science Council of Taiwan (NSC-100-2917-I-007-006),
the Erasmus Mundus Programme of the European Union, and
the German Research Foundation,
respectively.

\end{document}